\documentclass[a4paper,twocolumn]{article}
\usepackage{amsmath}
\usepackage{amsthm}
\newtheorem{theorem}{Theorem}
\newtheorem{proposition}[theorem]{Proposition}
\newtheorem{corollary}[theorem]{Corollary}
\newtheorem{lemma}[theorem]{Lemma}
\theoremstyle{remark}
\newtheorem{example}[theorem]{Example}
\newcommand{\while}{\mathit{while}}
\newcommand{\univ}{\mathbf{dom}}
\newcommand{\adom}[1]{\mathrm{adom}(#1)}
\newcommand{\Schtime}{\Sch^{\rm time}}
\newcommand{\Tape}{{\it Tape}}
\newcommand{\Begin}{{\it Begin}}
\newcommand{\End}{{\it End}}

\newcommand{\Lang}{\mathcal L}
\newcommand{\Sch}{\mathcal{S}}

\newcommand{\out}{\mathit{out}}
\newcommand{\Schin}{\Sch_{\mathrm{in}}}
\newcommand{\Schsys}{\Sch_{\mathrm{sys}}}
\newcommand{\Schmsg}{\Sch_{\mathrm{msg}}}
\newcommand{\Schmem}{\Sch_{\mathrm{mem}}}
\newcommand{\Qconstr}[3]{#1_{#2}^{#3}}
\newcommand{\Qsnd}[1]{\Qconstr Q{\mathrm{snd}}{#1}}
\newcommand{\Qins}[1]{\Qconstr Q{\mathrm{ins}}{#1}}
\newcommand{\Qdel}[1]{\Qconstr Q{\mathrm{del}}{#1}}
\newcommand{\Qout}{Q_{\mathrm{out}}}
\newcommand{\Trans}{\Pi}
\newcommand{\transi}[5]{#1,#2 \xrightarrow{#4} #5,#3}
\newcommand{\Ircv}{I_{\mathrm{rcv}}}

\newcommand{\Jsnd}{J_{\mathrm{snd}}}

\newcommand{\Jout}{J_{\mathrm{out}}}
\newcommand{\Net}{\mathcal N}
\newcommand{\Id}{\mathit{Id}}
\newcommand{\All}{\mathit{All}}

\newcommand{\conf}{\gamma}
\newcommand{\state}{\mathit{state}}
\newcommand{\buf}{\mathit{buf}}
\newcommand{\transn}[5]{#1 \xrightarrow[#2,#3]{#4} #5}
\begin{document}
\title{Relational transducers for declarative networking}
\author{Tom Ameloot\thanks{PhD Fellow of the Fund for
Scientific Research, Flanders (FWO).}, Frank Neven and Jan Van den Bussche \\
Hasselt University, Belgium \&
transnational University of Limburg}
\date{}
\maketitle
\begin{abstract}
Motivated by a recent conjecture concerning
the expressiveness of declarative networking, we propose a formal
computation model for ``eventually consistent'' distributed querying,
based on relational transducers.  A tight link has been
conjectured between coordination-freeness of computations, and
monotonicity of the queries expressed by
such computations.  Indeed, we propose a formal definition of
coordination-freeness and confirm that the class of 
monotone queries is captured by coordination-free transducer
networks.  Coordination-freeness is a semantic property, but the
syntactic class that we define of ``oblivious'' transducers also
captures the same class of monotone queries.
Transducer networks that are not coordination-free are much more
powerful.
\end{abstract}
\section{Introduction}

Declarative networking \cite{decl_netw_cacm} is a recent approach
by which distributed computations and networking protocols are
modeled and programmed using formalisms based on Datalog.  In his
keynote speech at PODS 2010
\cite{hellerstein_datalog,hellerstein_declimp}, Hellerstein made
a number of intriguing conjectures concerning the expressiveness
of declarative networking.  In the present paper we are focusing
on the CALM conjecture (Consistency And Logical Monotonicity).
This conjecture suggests a strong link between, on
the one hand, ``eventually consistent'' and ``coordination-free''
distributed computations, and on the other hand, expressibility
in monotonic Datalog (Datalog without negation or aggregate
functions).  The conjecture was not fully
formalized, however; indeed, as Hellerstein notes himself, a proper
treatment of this conjecture requires crisp definitions of
eventual consistency and coordination, which have been lacking so
far.  Moreover, it also requires a formal model of distributed
computation.

In the present paper, we investigate the CALM conjecture in
the context of a model for distributed database querying.  In the
model we propose, the computation is performed on a network of
relational transducers.  The relational transducer model,
introduced by Abiteboul and Vianu, is well established in
database theory research as a model for data-centric agents
reacting to inputs
\cite{av_reltrans,spielmann_reltrans,deutsch_website,deutsch_webservice}.
Relational transducers are firmly grounded in the theory of
database queries \cite{av_generic,av_fo} and also have close
connections with Abstract State Machines \cite{asm_ccql}.  It
thus seems natural to consider networks of relational
transducers, as we will do here.  We give a formal operational
semantics for such networks, formally define ``eventual
consistency'', and formally define what it means for a network to
compute a conventional database query, in order to address the
expressiveness issues raised by Hellerstein.

It is less clear, however, how to formalize the intuitive notion
of ``coordination''.  We do not claim to resolve this issue
definitively, but we propose a new, nonobvious definition that
appears workable.  Distributed algorithms requiring coordination
are viewed as less efficient than coordination-free algorithms.
Hellerstein has identified \emph{monotonicity} as a fundamental
property connected with coordination-freeness.  Indeed,
monotonicity enables ``embarrassing parallelism''
\cite{hellerstein_declimp}: agents working
on parts of the data in parallel can produce parts of the output
independently, without the need for coordination.

One side of the CALM conjecture now states that any database
query expressible in monotonic Datalog can be computed in a
distributed setting in an eventually consistent,
coordination-free manner.  This is the easy side of the
conjecture, and indeed we formally confirm it in the following
broader sense: any monotone query $Q$ can be computed by a
network of ``oblivious'' transducers.  Oblivious transducers are
unaware of the network extent (in a sense that we will make
precise), and every oblivious transducer network is
coordination-free.  Here, we should note that the transducer
model is parameterized by the query language $\mathcal{L}$ that
the transducer can use to update its local state.  Formally, the
monotone query $Q$ to be computed must be expressible in the
while-closure of $\mathcal{L}$ for the above confirmation to
hold.  If $Q$ is in Datalog, for example, then $\mathcal{L}$ can
just be the conjunctive queries.

The other side of the CALM conjecture, that the query computed by
an eventually consistent, coordination-free distributed program
is always expressible in Datalog, is false when taken literally,
as we will point out.
Nevertheless, we do give a Datalog version
of the conjecture that holds.  More importantly, we confirm the
conjecture in the following more general form: coordination-free networks
of transducers can compute only monotone queries.
Note that here we are using our newly proposed
formal definition of coordination-free.

Finally, the present work also lead us to think
about the computational power of the language Dedalus
\cite{dedalus}, the Datalog extension used by Hellerstein et al.
We will show that this language is quite powerful by establishing
a monotone simulation of arbitrary Turing machines.

This paper is organized as follows.  Preliminaries are in
Section~\ref{prelim}.  Section~\ref{secdef} introduces networks
of transducers.  Section~\ref{seccons} investigates the use of
transducer networks for expressing conventional database queries
in a distributed fashion.  Section~\ref{coordination} discusses
the issue of coordination.  Section~\ref{secalm} looks into the
CALM conjecture. Section~\ref{further} presents some further
results.  Section~\ref{secdedalus} compares our results to
the language Dedalus.  Section~\ref{secconcl} is the conclusion.

In this extended abstract, proofs are mainly given on an informal
level.

\section{Preliminaries} \label{prelim}

We recall some basic notions from the theory of database queries
\cite{ahv_book}.

A \emph{database schema} is a finite set $\Sch$ of relation names, each
with an associated arity (a natural number).  Assume some
infinite universe $\univ$ of atomic \emph{data elements}.  An
\emph{instance} of a database schema $\Sch$ is an assignment $I$ of
finite relations on $\univ$ to the relation names of $\Sch$,
such that when $R$ has arity $k$ then $I(R)$ is a $k$-ary
relation.  Equivalently, we can view an instance as a set of
\emph{facts} over $\Sch$, where a fact is an expression of the
form $R(a_1,\dots,a_k)$ with $a_1,\dots,a_k \in \univ$ and $R \in
\Sch$ of arity $k$.  The \emph{active domain} of an instance $I$,
denoted by $\adom I$, is the set of data elements occurring in $I$.

A \emph{$k$-ary query over $\Sch$} is a partial function $Q$
mapping instances of $\Sch$ to $k$-ary relations on $\univ$ such that 
for each $I$ on which $Q$ is defined, the following two
conditions hold:
\emph{(i)} $Q(I)$ is a $k$-ary
relation on $\adom I$; and \emph{(ii)} $Q$ is also defined
on the isomorphic instance $h(I)$, for each permutation $h$ of $\univ$,
and $Q(h(I)) = h(Q(I))$.  A query $Q$ is \emph{monotone} if for
any two instances $I \subseteq J$, if $Q(I)$ is defined then so
is $Q(J)$, and $Q(I) \subseteq Q(J)$.

We assume familiarity with \emph{first-order logic (FO)} as a
basic database query language.  An FO formula
$\varphi(x_1,\dots,x_k)$ expresses the $k$-ary query defined by
$\varphi(I) = \{(a_1,\dots,a_k) \in \adom I^k \mid (\adom I,I)
\models \varphi[a_1,\dots,a_k]\}$.  Note that we evaluate FO
formulas on instances under the active-domain semantics.  The
resulting query language is equivalent in expressive power to the
relational algebra, as well as to recursion-free Datalog with
negation.

We will also consider the query languages Datalog, stratified
Datalog (with negation), and \textit{while}.
Datalog and stratified Datalog are well known; \textit{while} is the
query language obtained from FO by adding assignment statements
and while-loops.
Finally, we recall that there exist quite elegant \emph{computationally
complete} query languages in which every partial computable query
can be expressed.

\subsection{Relational transducers}

A \emph{transducer schema} is a tuple
$(\Schin,\allowbreak\Schsys,\allowbreak\Schmsg,\allowbreak
\Schmem,\allowbreak k)$ consisting of four
disjoint database schemas and an \emph{output arity} $k$.
Here, `in' stands for `input'; `sys' stands for
`system'; `msg' stands for `message'; and `mem' stands for
`memory'.

An \emph{abstract relational transducer} (or just
transducer for short) over this schema is a collection of queries
$\{\Qsnd R \mid R \in \Schmsg\} \cup \{\Qins R \mid R \in
\Schmem\} \cup \{\Qdel R \mid R \in \Schmem\} \cup \{\Qout\}$,
where
\begin{itemize}
\item
every query is over the combined database schema $\Schin
\cup \Schsys \cup \Schmsg \cup \Schmem$;
\item
the arity of each
$\Qsnd R$, each $\Qins R$ and each $\Qdel R$ equals the arity of
$R$; and
\item
the arity of $\Qout$ equals the output arity $k$.
\end{itemize}
Here, `snd' stands for `send'; `ins' stands for `insert';
`del' stands for `delete'; and
`out' stands for `output'.

A \emph{state} of the transducer is an instance of the combined
schema $\Schin \cup \Schsys \cup \Schmem$.
Intuitively, a state just consists of some input relations, some
system relations (we will make these precise in the next
section), and some stored relations that constitute the memory of
the transducer.

A \emph{message instance} is, plainly, an instance of $\Schmsg$.
Such a message instance can stand for a set of messages (facts)
received by the transducer, but can as well stand for a set of
messages sent by the transducer.  It will always be clear from
the context which of the two meanings we have.

Let $\Trans$ be a transducer.
A \emph{transition} of $\Trans$ is a five-tuple
$(I,\Ircv,\Jsnd,\Jout,J)$, also denoted as $\transi I \Ircv \Jsnd
\Jout J$,
where $I$ and $J$ are states, $\Ircv$ and $\Jsnd$ are message
instances, and $\Jout$ is a $k$-ary relation such that
\begin{itemize}
\item
every query of $\Trans$ is defined on $I' = I \cup \Ircv$;
\item
$J$ agrees with $I$ on $\Schin$ and $\Schsys$;
\item
$\Jsnd(R)$, for each $R \in \Schmsg$, equals $\Qsnd R(I')$;
\item
$\Jout$ equals $\Qout(I')$; 
\item
$J(R)$, for each $R \in \Schmem$, equals
\begin{multline*}
(\Qins R(I') \setminus \Qdel R(I')) \\ {} \cup
(\Qins R(I') \cap \Qdel R(I') \cap I(R)) \\ {} \cup
(I(R) \setminus (\Qins R(I') \cup \Qdel R(I'))).
\end{multline*}
\end{itemize}
The intuition behind the instance $I'$
is that $\Trans$ sees its input, system and memory relations, plus its
received messages.  The transducer does not modify the
input and system relations.  The transducer computes new
tuples that can be sent out as messages; this is the instance $\Jsnd$.
The transducer also outputs some tuples; this is the relation
$\Jout$.  These outputs cannot
later be retracted!
Finally the transducer updates its memory by inserting and
deleting some tuples for every memory relation.  The intimidating
update formula merely expresses that conflicting inserts/deletes are ignored
\cite{spielmann_reltrans,deutsch_website}.
Note that an assignment $R := Q$ can be expressed by using $Q$ for
$\Qins R$ and $R$ for $\Qdel R$.

Note also that transitions are \emph{deterministic:}
if $\transi I \Ircv \Jsnd \Jout J$ and $\transi
I \Ircv {\Jsnd'} {\Jout'} {J'}$, then $\Jsnd'=\Jsnd$;
$\Jout'=\Jout$; and $J'=J$.

An abstract relational transducer as defined above is just a
collection of queries.  If we want to write down a transducer
then we will of course use some query language to express these
queries.  By default, we use FO as the query language.  More
generally, for any query language $\Lang$ we can consider the
language of \emph{$\Lang$-transducers} consisting of all transducers
whose queries are expressed in $\Lang$.  Because we are going to
place transducers in networks, we can think of $\Lang$ as the
language that individual peers use locally.  For
example, in the language Dedalus \cite{dedalus}, the local
language is stratified Datalog.

\section{Transducer networks} \label{secdef}

\newtheorem*{proviso}{Proviso}
\begin{proviso}
From now on we will only consider transducers where the system
schema $\Schsys$ consists of the two unary relation names $\Id$ and $\All$.
\end{proviso}

A \emph{network} is a finite, \emph{connected}, undirected graph over a
set of vertices $V \subset \univ$.  We refer to the vertices as 
\emph{nodes}.  Note that nodes belong to the universe $\univ$ of
atomic data elements; indeed
we are going to allow that nodes are stored in relations.
We stress again that a network must be connected.  This is
important to make it possible for flow of information to
reach every node.

A \emph{transducer network} is a pair $(\Net,\Trans)$ where
$\Net$ is a network and $\Trans$ is a transducer.
The general idea is that a copy of $\Trans$ is running
on every node.  A database instance is distributed over the input
relations of the different nodes.  Relation $\Id$ contains the
node identifier where the transducer is running, and relation
$\All$ is the same at all nodes and contains the set of all
nodes.  When a node $v$ sends a set of
facts as messages, these facts are added to the message buffers
of $v$'s neighbors.  Nodes receive facts one by one in arbitrary
order; messages are not necessarily received in the order they
have been sent.  A similar situation can happen in the Internet
with subsequent TCP/IP connections between the same two nodes,
where an earlier connection might be slower than a later one.
Moreover, nodes
regularly receive a ``heartbeat'' message which allows them to
transition even when no message is read.

We proceed to define the possible runs of a transducer network more formally.
A \emph{configuration} of the system is a pair $\conf=(\state,\buf)$ of
mappings where
\begin{itemize}
\item
$\state$ maps every node $v$ to a state $I = \state(v)$ of
$\Trans$, so that 
$I(\Id) = \{v\}$, and $I(\All) = V$ (the set of all nodes of
$\Net$).
\item
$\buf$ maps every node to a finite multiset of facts over $\Schmsg$.
\end{itemize}
Thus, the system relations $\Id$ and $\All$ give the transducer
knowledge about the node where it is running and about the other
nodes in the network.  We will discuss the use and necessity of
these relations extensively.

A \emph{general transition} of the system is the transformation
of one configuration to another where some node
$v$ reads and removes some message instance
$\Ircv$ from its input buffer, makes a local
transition, and sends the resulting message instance $\Jsnd$ to its neighbors.
Formally, a general transition
is a tuple $\tau=(\conf_1,v,\Ircv,\Jout,\conf_2)$,
also denoted as $\transn {\conf_1} v \Ircv \Jout {\conf_2}$, where
$\conf_i=(\state_i,\buf_i)$ for $i=1,2$ is a configuration,
$v$ is a node, and $\Ircv \subseteq \buf_1(v)$ (multiset
containment), such that:
\begin{itemize}
\item
$\state_2(v') = \state_1(v')$ for every node $v' \neq v$.
\item
There exists $\Jsnd$ such that
$\transi {\state_1(v)} \Ircv \Jsnd \Jout {\state_2(v)}$. We
call $\Jout$ the \emph{output} of the transition and denote it
also by $\out(\tau)$.  Note that, since individual transducer transitions
are deterministic, $\Jout$ and $\Jsnd$ are uniquely determined by
$\state_1(v)$ and $\Ircv$.
\item
$\buf_2(v) = \buf_1(v) \setminus \Ircv$ (multiset difference).
\item
$\buf_2(v') = \buf_1(v')$ for every node $v'\neq v$ that is not a neighbor
of $v$.
\item
$\buf_2(v') = \buf_1(v') \cup \Jsnd$ for every node $v'$ that is
a neighbor of $v$.  Note we are using multiset union here.
\end{itemize}

We will, however, not use transitions in their most general form but
only in two special forms:
\begin{description}
\item[Heartbeat transition:] is of the form
$\transn {\conf_1} v \emptyset \Jout {\conf_2}$.  So, some node $v$
transitions without reading any message.
\item[Delivery transition:] is of the form
$\transn {\conf_1} v {\{f\}} \Jout {\conf_2}$.  So, some node $v$
reads a single fact $f$ from its received message buffer.
\end{description}
We only consider these two forms because they appear to be the
most elementary.  We are not sure it is realistic to assume that
entire message instances can be read in one transition.
Therefore we limit message reading to a single fact.  Heartbeat
transitions ensure that nodes can transition even if their
message buffer is empty.

For any two configurations $\conf_1$ and $\conf_2$ we simply
write $\conf_1 \to \conf_2$ to denote that the system can
transition from $\conf_1$ to $\conf_2$ either by some heartbeat
transition or by some delivery transition.
A \emph{run} of the system now is an infinite sequence $(\tau_n)_n$ 
of heartbeat or delivery transitions such that
for each $n$, if
$\tau_n$ is $\conf_n \to \conf_{n+1}$ then $\tau_{n+1}$ is of
the form $\conf_{n+1} \to \conf_{n+2}$.  In other words, each
transition $\tau_n$ with $n>0$ works on the result configuration
of the previous transition.

The \emph{output} of a run $\rho$ is then
defined as $\out(\rho) = \bigcup_n \out(\tau_n)$.
We note the following:
\begin{proposition} \label{quiescence}
For every run $\rho = (\tau_n)_n$ there exists a natural number
$m$ such that $\out(\rho) = \bigcup_{n=0}^m \out(\tau_n)$.
The number $m$ is called a \emph{quiescence point} for $\rho$.
\end{proposition}
Indeed, since the initial configuration contains only a finite
number of distinct atomic data elements, and a local query cannot
invent new data elements, only a finite number of distinct
output tuples are possible.

In the language Dedalus \cite{dedalus}, new data elements are
invented in the form of increasing timestamps, so the above
proposition does not hold for Dedalus.  It would be interesting
to investigate a version of transducer networks with timestamps,
or with object-creating local queries.

\section{Expressing queries with transducer networks}
\label{seccons}

What does it take for a transducer network to compute some global
query?  Here we propose a formal definition based on the two
properties of \emph{consistency} and \emph{network-topology independence}.

An instance $I$ of the input part of the transducer schema,
$\Schin$, can be distributed over the input
relations of the nodes on the network.  Formally, for any
instance $I$ of $\Schin$, a \emph{horizontal partition of $I$ on
the network $\Net$} is a function $H$ that maps every node $v$
to a subset of $I$, such that $I = \bigcup_v H(v)$.
The \emph{initial configuration for $H$}
is a configuration $(\state,\buf)$ such that:
\begin{itemize}
\item
$\buf(v)=\emptyset$
and $\state(v)(R)=\emptyset$ for each node $v$ and every $R \in
\Schmem$. So, each node starts with an empty message buffer and
empty memory;
\item
For each node $v$, the restriction of $\state(v)$ to
the input schema $\Schin$ equals $H(v)$.
\end{itemize}
A \emph{run on $H$} is a run that starts in the initial
configuration for $H$.

We also need the notion of \emph{fair} run.  A run is fair
if every node does a heartbeat transition infinitely often, and
every fact in every message buffer is eventually taken out by a
delivery transition.  We omit the obvious formalization.

We now say that a transducer network $(\Net,\Trans)$ is
\emph{consistent} if for every instance $I$ of $\Schin$,
all fair runs on all possible horizontal partitions of $I$
have the same output.  Naturally, a consistent transducer
network is said to \emph{compute}
a query $Q$ over $\Schin$ if for every
instance $I$ of $\Schin$ on which $Q$ is defined, every fair run
on any horizontal partition of $I$ outputs $Q(I)$.

\begin{example}
Let us see a simple example of a network that is \emph{not} consistent.
Consider a network with
at least two nodes (indeed if the network consists of a
single node the transducer runs
all by itself; no messages are delivered
and there is only one possible run).
The input is a set $S$ of data elements.  Each node sends its
part of $S$ to its neighbors.  Also, each node outputs the first
element it receives and outputs no further elements.
This is easily programmed on an FO-transducer.  When there are at
least two nodes and at least two elements in $S$, different
runs may deliver the elements in different orders, so different
outputs can be produced, even for the same horizontal partition. \qed
\end{example}

\begin{example} \label{tc}
For a simple example of a consistent network, let the input be a
binary relation $S$.  Each node outputs the identical pairs from its
part of the input.  No messages are sent.  This network computes
the equality selection $\sigma_{\$1=\$2}(S)$.

An example of a consistent transducer
network that involves communication, computes the
transitive closure of $S$ in a distributed fashion in the
well-known way \cite{decl_netw_cacm}.  We present here, a naive,
unoptimized version.  Each node sends its part
of the input to its neighbors.  Each node also sends all tuples
it receives to its neighbors.  In this way the input is flooded
to all nodes.  Each node accumulates the tuples it receives in a
memory relation $R$.  Finally, each node maintains a memory
relation $T$ in which we repeatedly insert $S \cup R \cup T \cup (T
\circ T)$ (here $\circ$ stands for relational composition).  This
relation $T$ is also output.  Thanks to the monotonicity of the
transitive closure, this transducer network is consistent.
\qed
\end{example}

We are mainly interested in the case where the query can be
correctly computed by the distributed transducer program,
regardless of the network topology.  For example, the transitive
closure computation from Example~\ref{tc} is independent of the
network topology (as long as the network is connected, but we
are requiring that of all networks).

Formally, a transducer $\Trans$ is \emph{network-topology
independent} if for every network $\Net$, the system
$(\Net,\Trans)$
is consistent, and regardless of $\Net$ computes the same query
$Q$.  We say that $Q$ can be \emph{distributedly computed} by
$\Trans$.

\begin{example}
For a simple example of a transducer that is consistent for every
network topology, but that is not network-topology independent,
consider again as input a set $S$ distributed over the nodes.
Each node sends its input to its neighbors and also sends the
elements it receives to its neighbors.  Each node only outputs
the elements it receives.  On any network with at least two
nodes, the identity query is computed, but on the network with a
single node, the empty query is computed.
\qed
\end{example}

In order to state a few first results in
Theorem~\ref{observations}, we introduce the following terminology.
\begin{description}
\item[Oblivious transducer:] does not use the relations $\Id$ and
$\All$.  Intuitively, the transducer program is unaware of the
context in which it is running.
\item[Inflationary transducer:] does not do deletions, i.e., each
deletion query returns empty on all inputs.
\item[Monotone transducer:] uses only monotone local queries.
\end{description}

\begin{lemma} \label{flood}
\begin{enumerate}
\item
There is an inflationary FO-transducer such that, on any network,
starting on any horizontal partition of any instance $I$ of
$\Schin$, any fair run reaches a configuration where every node
has a local copy of the entire instance $I$ in its memory, and an
additional flag \textit{Ready} (implemented by a nullary memory
relation) is true.  Moreover, the flag \textit{Ready} does not
become true at a node before that node has the entire instance in
its memory.
\item
There is an oblivious, inflationary, monotone FO-transducer that
accomplishes the same as the previous one, except for the flag
\textit{Ready}.
\item
A query is expressible in the language `$\while$' if and only if it is
computable by an FO-transducer on a single-node network.
\end{enumerate}
\end{lemma}
\begin{proof}
For (1), a multicast protocol \cite{attiyawelch_dcbook} is implemented.
Every node $v$ sends out all the facts in its local input
relations, but each fact is tagged with the id of $v$ in an extra
coordinate (using relation $\Id$).  Every
node also forwards all input facts it receives,
and stores received facts in memory.  
Moreover, for every input fact received, every node sends an
acknowledge fact, additionally tagged with its own identifier.
Every node $v$ keeps a record of for which of its local input facts it has
received an ack from which node.  When $v$ has received an ack
from $v'$ for every local input fact,
it sends out a message $\mathit{done}(v,v')$ meant for $v'$.
When a node has received
\textit{done} from every node
(which can be checked using the relation $\All$),
it knows it is ready.
No deletions are necessary.

For (2), the program is much simpler.  All nodes simply send out
their local input facts and forward any message they receive.  In
any fair run, eventually all nodes will have received all input
facts.  Relations $\Id$ and $\All$ are not needed.

For (3),
on a one-node network there are only heartbeat transitions.
A $\while$ program can be simulated
by iterated heartbeats using well-known techniques
\cite{av_datalog}.  Conversely, it is clear that a one-node
transducer network can be simulated by a $\while$ program.
The only difficulty is that the transducer keeps running
indefinitely whereas the $\while$ program is supposed to stop.
Using the technique described by Abiteboul and Simon
\cite{as_datalog}, however, the program can detect that it is in
an infinite loop.  This implies that the transducer has repeated
a state and will output no more new tuples.
\end{proof}

\begin{theorem} \label{observations}
\begin{enumerate}
\item \label{everyquery}
Every query can be distributedly computed by some abstract
transducer.  In particular, if $\Lang$ is computationally
complete, every partial computable query can be distributedly computed
by an $\Lang$-transducer.
\item \label{everymonotone}
Every monotone query can be 
distributedly computed by an oblivious, inflationary, monotone
abstract transducer.
In particular, if $\Lang$ is computationally complete, every
partial computable monotone query can be distributedly computed
by an oblivious $\Lang$-transducer.
\item \label{whileresult}
A query is expressible in the language `$\while$'
if and only if it can be distributedly
computed by an FO-transducer.
\item \label{monotonewhile}
Every \emph{monotone} query expressible in
$\while$ can be distributedly computed by an oblivious
FO-transducer.
\item \label{datalogresult}
A query is in Datalog if and only if it can be
distributedly computed by an
oblivious, inflationary, nonrecursive-Datalog-transducer.
\end{enumerate}
\end{theorem}
\begin{proof}
For (\ref{everyquery}), to distributedly compute any query $Q$, we first run the
transducer from Lemma~\ref{flood}(1) to obtain the entire
input instance. Then we apply and output $Q$.

For (\ref{everymonotone}),
the idea is the same, but we now use the transducer from
Lemma~\ref{flood}(2).  We continuously apply $Q$
to the part of the input instance already received, and output
the result.  Since $Q$ is monotone, no incorrect tuples are
output.

For (\ref{whileresult})
we only still have to argue the only-if implication.  We
first run the transducer from
Lemma~\ref{flood}(1) to obtain the entire input instance.
Then every node can act as if it is on its own, ignoring any
remaining incoming messages and simulate the $\while$-program.

For (\ref{monotonewhile}) the idea is the same as in 
(\ref{everymonotone}).  We receive input tuples and store them in
memory.  We continuously recompute the $\while$-program, starting
afresh every time a new input fact comes in.  We use
deletion to start afresh.  Since the query is monotone, no
incorrect tuples are output.

For (\ref{datalogresult}) the idea for the only-if implication
is again the same as in
(\ref{everymonotone}).  We receive input tuples
and apply continuously the $T_P$-operator of the Datalog program.
By the monotone nature of Datalog evaluation,
deletions are not needed, so the transducer is inflationary.
The Datalog program for the if-implication is obtained by taking
together the rules of all update queries $\Qins R$ and the output
query $\Qout$.  The answer predicate of $\Qout$ is the global
answer predicate.  The answer predicate of each $\Qins R$ is
replaced by $R$, so that we obtain a recursive program.
\end{proof}

Without proof we note that the transducer from
Lemma~\ref{flood}(1)
can actually be implemented to use only unions of
conjunctive queries with negation (UCQ$\neg$).
By simulating FO queries by fixed compositions of UCQ$\neg$, we
obtain: (proof omitted)
\begin{proposition}
Every (monotone) query that can be distributedly computed by an
FO-transducer can be distributedly computed by an (oblivious)
UCQ$\neg$-transducer.
\end{proposition}

To conclude, we remark that in a transducer network of at least two
nodes, each node can establish a linear order on the active
domain, by first collecting all input tuples, then sending out
all elements of the active domain, forwarding messages and
storing the elements that are received back in the order they are
received.
But such a transducer is not truly
network-topology independent, as it does not work in the same way
on a one-node network.  At any rate, by the well-known
characterization of PSPACE \cite{ahv_book}, we obtain:
\begin{corollary}
On any network with at least two nodes,
every PSPACE query can be computed by an FO-transducer.
\end{corollary}

\section{Coordination} \label{coordination}

The CALM conjecture hinges on an intuitive notion of
``coordination-freeness'' of certain distributed computations.
For some tasks, coordination is required to reach consistency
across the network.  Two-phase commit is a classical example.
The multicast protocol used in Lemma~\ref{flood}(1) also requires
heavy coordination.  Since coordination typically
blocks distributed computations, it is good to understand
precisely when it can be avoided.  This is what the CALM
conjecture is all about.

It seems hard to give a definitive formalization of coordination.
Still we offer here a nontrivial definition that appears interesting.
A very drastic, too drastic, definition of coordination-free would be to
disallow any communication.  Our definition is much less severe
and only requires that the computation can succeed without
communication on ``suitable'' horizontal partitions.  It actually does
not matter what a suitable partition is, as long as it
exists.  Even under this liberal definition, the CALM conjecture
will turn out to hold.

Formally, consider a network-topology independent transducer
$\Trans$ and a network $\Net$.  We call $\Trans$
\emph{coordination-free on $\Net$} if for every instance $I$ of
$\Schin$, there exists a horizontal partition $H$ of $I$ on
$\Net$ and a run $\rho$ of $(\Net,\Trans)$ on $H$, in which a
quiescence point (Proposition~\ref{quiescence}) is already
reached by only performing heartbeat transitions.  Intuitively,
if the horizontal partition is ``right'', then no communication
is required to correctly compute the query.  Finally we call
$\Trans$ \emph{coordination-free} if it is coordination-free on
any network.

\begin{example}
Consider again the transitive closure computation from
Example~\ref{tc}.  When every node already has the full input,
they can each individually compute the transitive closure.  Hence
this transducer is coordination-free.

The reader should not be lulled into believing that with a
coordination-free program it is always
sufficient to give the full input at all nodes.  A (contrived)
example of a coordination-free transducer that needs communication even if
each node has the entire input is the following.  The input,
distributed over the nodes as usual, consists of two sets $A$ and
$B$, and the query is to determine if at least one of $A$ and $B$
are nonempty.  If the network has only one node (which can be
tested by looking at the $\All$ relation), the transducer simply
outputs the answer to the query.  Otherwise, it
first tests if its local
input fragments $A$ and $B$ are both nonempty.  If yes,
nothing is output, but the value `true' (encoded by the empty
tuple) is sent out.  Any node that receives the message `true'
will output it.  When $A$ or $B$ is empty locally,
the transducer simply outputs the desired output directly.
The transducer is coordination-free, since if we take care to
have at least one node that knows $A$, and another node that
knows $B$, but no node knows both, then the query can be computed
without communication.  When
$A$ and $B$ are both nonempty, however, a run on the horizontal
partition where every node has the entire input will not reach
quiescence without communication.  \qed
\end{example}

\begin{example}
A simple example of a transducer that is not coordination-free,
i.e., requires communication, is the one that computes the
emptiness query on an input set $S$.
Since every node can have a part of the
input, the nodes must coordinate with each other to be certain
that $S$ is empty at every node.  Every node sends out its
identifier (using the relation $\Id$)
on condition that its local relation $S$ is empty.
Received messages are forwarded, so that if $S$ is globally
empty, eventually all nodes will have received the identifiers of
\emph{all} nodes, which can be checked using the relation $\All$.
When this has happened the transducer at each node outputs `true'.
\qed
\end{example}

Coordination-freeness is undecidable for FO-transducers, but we
can identify a syntactic class of transducers that is guaranteed
to be coordination-free, and that will prove to have the same
expressive power as the class of coordination-free transducers.
Specifically, recall that an \emph{oblivious} transducer is one
that does not use the system relations $\Id$ and $\All$.
For now we observe:

\begin{proposition} \label{oblivious}
Every network-topology independent, oblivious transducer
is coordination-free.
\end{proposition}
\begin{proof}
Let $\Trans$ be a network-topology independent, oblivious transducer.
Let $Q$ be the query distributedly computed by $\Trans$.
On a one-node network and given any input instance $I$,
transducer $\Trans$
reaches quiescence and outputs $Q(I)$ by doing only heartbeat transitions.
Now consider any other network, any instance $I$ over $\Schin$,
and the horizontal partition $H$ that places the entire $I$ at
every node.  Since $\Trans$ is oblivious, nodes cannot detect
that they are on a network with multiple nodes unless they
communicate.  So, by doing only heartbeat transitions initially,
every node will act the same as if in a one-node network and will
already output the entire $Q(I)$.
\end{proof}

\section{The CALM conjecture} \label{secalm}

\theoremstyle{plain}
\newtheorem*{calm}{CALM Conjecture}
\begin{calm}[\cite{hellerstein_declimp}]
A program has an
eventually consistent, coordination-free execution strategy if
and only if it is expressible in (monotonic) Datalog.
\end{calm}

It is not specified what is meant by ``program'' or ``strategy'';
here, we will take these terms to mean ``query'' and
``distributedly computable by a transducer'',
respectively.  The term ``eventually consistent'' is
then formalized by our notions of consistency and
network-topology independence.
Under this interpretation, the conjecture becomes ``a
query can be distributedly computed by a coordination-free
transducer if and only if it is expressible in
Datalog''.

Let us immediately get the if-side of the conjecture out of
the way.  It surely holds, and many versions of it are already
contained in Theorem~\ref{observations}.  That theorem talks
about oblivious transducers, but we have seen in
Proposition~\ref{oblivious} that these are coordination-free.

As to the only-if side, the explicit mention of Datalog is a bit
of a nuisance.  Datalog is limited to polynomial time whereas there
certainly are monotone queries outside PTIME\@.  This continues
to hold for queries expressible in the language Dedalus that
Hellerstein uses; we will show this in Section~\ref{secdedalus}.
We also mention the celebrated paper \cite{acy_datalog} where
Afrati, Cosmadakis and Yannakakis show that even within PTIME
there exist queries that are monotone but not expressible in
Datalog.

We will see in Corollary~\ref{maincorol}(\ref{datalogcalm}) how a
Datalog version of the CALM conjecture can be obtained.
Datalog aside, however,
the true emphasis of the CALM conjecture clearly lies in the monotonicity
aspect. Indeed we confirm it in this sense:
\begin{theorem} \label{confirmation}
Every query that is distributedly computed by a coordination-free
transducer is monotone.
\end{theorem}
\begin{proof}
Let $Q$ be the query distributedly computed by the
coordination-free transducer $\Trans$.  Let $I \subseteq J$ be
two input instances and let $t \in Q(I)$.  We must show $t \in
Q(J)$.  Consider a network $\Net$ with at least two nodes.
Since $\Trans$ is coordination-free, there exists a horizontal
partition $H$ of $I$ such that $Q(I)$
is already output distributedly over the nodes, by letting the nodes do
only heartbeat transitions.  Let $v$ be a node where $t$ is
output.  Let $v'$ be a node different from $v$ and consider the
horizontal partition $H'$ of $J$ where $H'(v)=H(v)$ and $H'(v') =
H(v') \cup (J \setminus I)$.  The partial run on $H$ where $v$ first does
only heartbeat transitions until $t$ is output, is also a partial
run on $H'$.  This partial run can be extended to a fair run,
so $t$ is output by some fair run of $(\Net,\Trans)$ on $H'$.
Since $(\Net,\Trans)$ is consistent, $t$ will also be
output in any other fair run on any horizontal partition of $J$.
Hence, $t$ belongs to the query
computed by $(\Net,\Trans)$ applied to $J$. Moreover, $\Trans$ is
network-topology independent, so $t$ belongs to $Q(J)$.
\end{proof}

\begin{corollary}[CALM Property] \label{calmproperty}
The following are equivalent for any query $Q$:
\begin{enumerate}
\item
$Q$ can be distributedly computed by a transducer that is
coordination-free.
\item
$Q$ can be distributedly computed by a transducer that is
oblivious.
\item
$Q$ is monotone.
\end{enumerate}
\end{corollary}
\begin{proof}
Theorem~\ref{observations} yields $(3)\Rightarrow(2)$;
Proposition~\ref{oblivious} yields $(2)\Rightarrow(1)$;
Theorem~\ref{confirmation} yields $(1)\Rightarrow(3)$.
\end{proof}

Similarly we obtain the following versions of the CALM property:

\begin{corollary} \label{maincorol}
The following groups of statements are equivalent for any query $Q$:
\begin{enumerate}
\item Let $\Lang$ be a computationally complete query language.
\begin{enumerate}
\item
$Q$ can be distributedly computed by a coordination-free $\Lang$-transducer.
\item
$Q$ can be distributedly computed by an oblivious $\Lang$-transducer.
\item
$Q$ is monotone and partial computable.
\end{enumerate}
\item
\begin{enumerate}
\item
$Q$ can be distributedly computed by a coordination-free FO-transducer.
\item
$Q$ can be distributedly computed by an oblivious FO-transducer.
\item
$Q$ is monotone and expressible in the language `$\it while$'.
\end{enumerate}
\item \label{datalogcalm}
\begin{enumerate}
\item
$Q$ can be distributedly computed by a coordination-free
nonrecursive-Datalog-transducer.
\item
$Q$ can be distributedly computed by a
nonrecursive-Datalog-transducer that is oblivious.
\item
$Q$ is expressible in Datalog.
\end{enumerate}
\end{enumerate}
\end{corollary}

\section{Further results} \label{further}

It is natural to wonder about
variations of our model.  One question may be
about the system relations $\Id$ and $\All$.  Without them (the
oblivious case) we know that we are always coordination-free and
thus monotone.  What if we have only $\Id$ or only $\All$?
As to coordination-freeness, it is readily verified that the
argument given in the proof of Proposition~\ref{oblivious} still
works in the presence of $\Id$.  It does not work in the presence
of $\All$, and indeed we have the following counterexample.
\begin{example}
We describe a transducer that is network-topology independent,
does not use $\Id$, but that is not coordination-free.  The query
expressed is simply the identity query on a set $S$.  The transducer
can detect whether he is alone in the network by looking at the
relation $\All$.  If so, he simply outputs the result.  If he is
not alone, he sends out a ping message.  Only upon receiving a
ping message he outputs the result.  Regardless of the horizontal
partition, on a multiple-node network, communication is required
for the transducer network to produce the required output.
\qed
\end{example}
So, coordination-freeness is not guaranteed when we use the
relation $\All$, but yet, monotonicity is not lost:
\begin{theorem}
Every query distributedly computed by a transducer that does not
use the system relation $\Id$, is monotone.
\end{theorem}
\begin{proof}
Let $\Trans$ be a network-topology independent transducer
and let $Q$ be the query distributedly computed by $\Trans$.
Let $I \subseteq J$ be two input instances and let $t \in Q(I)$.
We must show $t \in Q(J)$.  Consider the network $\mathcal R_4$
with four nodes 1--2--3--4--1 in the form of a ring.
Let $H$ be the horizontal partition of $I$ that places the entire
$I$ at every node.  Consider now the following, fair, run $\rho$ of
$(\mathcal R_4,\Trans)$ on $H$.  This particular run has
a fifo behavior of the message buffers.   We go around the network in
rounds.  The construction is such (proof omitted)
that after each round, all
nodes have the same state and the same fifo message buffer queue.
In each round, we first let each node do a heartbeat
transition.  Then, if some (hence every) input buffer is
nonempty, let each node do a delivery transition,
receiving the first tuple in its message buffer.
If the buffers are empty, we let each node do a second heartbeat
transition.
Since $t \in Q(I)$, we know that $t$ is output during run $\rho$.
Without loss of generality, assume node $1$ outputs tuple $t$
in round $m$ during run $\rho$.  

We now consider the modified network $\mathcal R'$ on the same
four nodes, obtained by adding the shortcut 2--4 to $\mathcal
R_4$.  Consider the horizontal partition $H'$ of $J$ defined by
$H'(1)=H'(2)=H'(4)=I$ and $H'(3)=J\setminus I$.  Consider now the
following prefix $\rho'$ of a possible run of $(\mathcal R',\Trans)$ on
$H'$.  The idea is that run $\rho$ is mimicked until round $m$,
but we ignore node 3 completely.  The construction is such
(proof omitted) that after each round, nodes 1, 2 and 4 have the same
state and the same fifo message buffer queue as after the same
round in $\rho$.  In each round $i$, we first let each node 1, 2 and
4 do a heartbeat transition.  Then, if in the same round in
  $\rho$ we made delivery transitions, then we make the same
  delivery transitions in $\rho'$ but not for node 3.  If in
  round $i$ we did a series of second heartbeat transitions,
  we do the same in $\rho'$ but again not for node 3.  

The result is that $t$ is also output by node $1$ during any fair run
that has $\rho'$ as a prefix.
Since $\Trans$ is network-topology independent, we
have $t \in Q(J)$ as desired.
\end{proof}

As a corollary we can add two more statements to the 
three equivalent
statements of the CALM Property (Corollary~\ref{calmproperty}):
\begin{corollary}
The following are equivalent for any query $Q$:
\begin{enumerate}
\item
$Q$ can be distributedly computed by an oblivious transducer.
\item
$Q$ can be distributedly computed by a transducer that does not
use the $\Id$ relation.
\item
$Q$ can be distributedly computed by a transducer that does not
use the $\All$ relation.
\end{enumerate}
\end{corollary}

To conclude this section we note that distributed algorithms
involving a form of coordination typically require the
participating nodes to have some knowledge about the other
participating nodes \cite{attiyawelch_dcbook}.  This justifies
our modeling of this knowledge in the form of the system
relations $\Id$ and $\All$.  Importantly, we have shown
that these relations are only necessary if one wants to compute
a nonmonotone query in a distributed fashion.

\section{Dedalus} \label{secdedalus}

Dedalus \cite{dedalus} is the declarative language used by
Hellerstein et al.\ to model and program network protocols.  The
precise expressive power of Dedalus needs to be better
understood.  Here, we compare Dedalus to our setting and we also
show that Dedalus can at least simulate
arbitrary Turing machines in an eventually consistent
manner.\footnote{This is not the same as saying that all
computable queries can be expressed in Dedalus; we conjecture
this is not the case.}  By the time hierarchy theorem
\cite{papadimitriou}, it follows that eventually-consistent
Dedalus programs are not contained in PTIME, let alone in
Datalog.

Dedalus is a temporal version of Datalog with negation where the
last position of each predicate carries a timestamp.  All
subgoals of any rule must be joined on this timestamp.  The timestamp
of the head of the rule can either be the timestamp of the body
(a ``deductive rule''), or it can be the successor timestamp (an
``inductive rule'').  The deductive rules must be stratifiable,
thus guaranteeing modular stratification and a
deterministic semantics through a unique minimal model.
Note how this corresponds well to
transducers using stratified Datalog as local query language.

Furthermore, Dedalus has a non-deterministic construct by which
facts can be derived with a random timestamp, used to model
asynchronous communication.  In our transducer networks, the same
effect is achieved by the semantics we have given, by which one
node may send a message in its $n$th local step, whereas another
node may receive the message in its $m$th local step where $m$
can be smaller as well as larger than $n$.  As long as a Dedalus
program is monotone in the relations derived by asynchronous
rules, the program remains deterministic, but there is no longer
a simple syntactic guarantee for this.

The feature that makes Dedalus quite powerful is that timestamp
values can also occur as data values, i.e., in other predicate
positions than the last one.  This feature, called
``entanglement'', is intriguing and makes Dedalus go beyond
languages such as temporal Datalog \cite{ci_infinite}.  Note that
entanglement does not involve arithmetic on timestamps; it merely
allows them to be copied in relations in a safe, Datalog-like
manner.

Turing machine simulations in database query languages are well
known \cite{av_datalog,ahv_book,immerman_book}, but the
Dedalus setting is new, so we describe the Turing
machine simulation in some detail.  For any database
schema $\Sch$ we can consider the database schema $\Schtime$ with
the same relation names as $\Sch$, but in $\Schtime$ each
relation name has arity one higher than in $\Sch$, in order to
accommodate timestamps.  Dedalus works with temporal database
instances; these are instances over schemas of the form
$\Schtime$ in which the last coordinate of every fact is a
natural number acting as timestamp.  For any instance $I$ over
$\Schtime$ and any timestamp value $n$, let $I|_n$ be the
instance over $\Sch$ obtained from the facts in $I$ that
have timestamp $n$, and let $\hat I$ equal $\bigcup_n I|_n$.

Now let $\Sigma$ be an arbitrary but fixed finite
alphabet, and consider the database schema $\Sch_\Sigma$ consisting
of relation names $\it Tape$ of arity two, $\it Begin$ and $\it End$ of arity
one, and $a$ of arity one for each $a\in\Sigma$.
Recall that any string $s = a_1\dots a_p$
over $\Sigma$ can be presented as an instance $I_s$ over
$\Sch_\Sigma$.  We consider only strings of
length at least two.  Then $I_s$ consists of the facts $\it
Tape(1,2)$, \dots, $\it Tape(p-1,p)$, $\it Begin(1)$, $\it End(p)$,
$a_1(1)$, \dots, $a_p(p)$.  Such instances, and isomorphic
instances, are known as \emph{word structures} \cite{thomas}.

For any Turing machine $M$, we define the boolean (0-ary) query $Q_M$ over the
class of temporal instances over $\Schtime_\Sigma$ as follows.
\begin{itemize}
\item
If $\hat I$ is a word structure
representing string $s$, and $M$ accepts $s$, then 
$Q_M(I)$ equals true (encoded by the 0-ary relation containing
the empty tuple).  If $M$ does not terminate on $s$, then
$Q_M(I)$ is undefined.
\item
If $\hat I$ contains a word structure,
but is not a word structure (due to spurious facts), then
$Q_M(I)$ also equals true.
\item
In all other cases $Q_M(I)$ equals false (encoded by the empty
0-ary relation).
\end{itemize}
The second item in the definition is there to ensure that $Q_M$
is monotone; nevertheless, when we give $Q_M$ a proper word structure as
input, a faithful simulation of $M$ is required.  Hence, the
computational complexity of $Q_M$ is as high as that of the
language accepted by $M$.

We say that a (deterministic) Dedalus program $\Pi$
\emph{expresses} a boolean query $Q$ over temporal instances, if for
every $I$ such that $Q(I)$ is defined, $\Pi(I)$ contains a fact
${\it Accept}(n)$ for some $n$ if and only if $Q(I)$ is true.
Moreover, $\Pi$ expresses $Q$ \emph{in an eventually consistent
way} if for every $I$ such that $Q(I)$ is defined, 
there exists $n$ such that $\Pi(I)|_m = \Pi(I)|_n$ for all $m \geq n$.

\begin{theorem}
  For every Turing machine $M$, the query $Q_M$ is expressible in an
  eventually consistent way by a Dedalus program.
\end{theorem}
\begin{proof}
We only sketch the proof and assume some familiarity with Dedalus.
The main difficulties to overcome are the following.
\begin{enumerate}
\item
Detection of a word structure.  Since input facts can arrive at
any timestamp, they are persisted, e.g.,
\begin{tabbing}
$a(x,n+1) \gets a(x,n)$ \qquad for each $a \in \Sigma$
\end{tabbing}
(Officially, this should be done using ``pos-predicates''
\cite{dedalus}.)
A word structure is detected at time $n$ if there is a
path in the $\Tape$ relation, beginning in an element in
$\Begin$, and ending in an element in $\End$, where all elements
on the path are \emph{labeled}, i.e.,
belong to some $a$ relation.  This is readily expressed in Datalog.
\item
Detection of spurious tuples.  When a word structure is already
detected, we can detect spurious tuples by checking for one of
the following conditions, which can be expressed in
stratified Datalog:
\begin{enumerate}
\item $\it Begin$ and $\it End$ contain more than a single element.
\item An element in the active domain is labeled by two different
alphabet letters.
\item $\it Tape$ is more than a successor relation from its begin
to its end point, i.e.,
there is an element on the tape with out-degree or in-degree more than one,
or there is an element on the tape that is not reachable from
$\it Begin$.
\item There exists a phantom element, i.e.,
an element in the active domain that is not labeled, or that is
not on the tape.
\end{enumerate}
\item
Turing machine simulation.
When a proper word structure is discovered, without spurious
tuples, the simulation of $M$ is started.  We copy the $a$ predicates to
$a^{\rm simul}$ predicates.  This is necessary because $a$ is
persisted, which would cause the simulation to be overwritten.
And we need to continue persisting $a$ 
because new (spurious) $a$ facts may
still arrive after the simulation has already started.
Each transition of $M$ goes to
the next timestamp.  For each state $q$ of $M$ we use
a predicate $q(x,n)$ that holds if $M$ at time $n$ is in state
$q$ with its head on position $x$.  Timestamp values
(entanglement) is used to extend the finite tape to the right.
Care must be taken to do this only when necessary, to ensure
eventual consistency.  Moreover, we must avoid confusing
timestamp values that may also already occur as input tape cells,
with timestamp values that are used to build the tape extension.
Thereto we use a separate predicate $\it TapeExt$ to represent
the tape extension.  For example, the first time that $M$ extends the
input tape, in some state $q$ and seeing letter $a$ at the last
input position, is expressed by the following rules:
\begin{tabbing}
${\it ExtNext}(x,n) \gets {\it TapeExt}(x,y,n)$ \\
${\it TapeExt}(x,n,n+1)
\gets q(x,n), a(x,n),$\\
\`$\End(x,n), \neg {\it ExtNext}(x,n)$
\end{tabbing}
For positions on the extension tape, we use predicates $q_{\rm
ext}$ instead of $q$ and $a^{\rm simul}_{\rm ext}$ instead of
$a^{\rm simul}$.
\end{enumerate}
\end{proof}

Distribution is not built in Dedalus and must be simulated using
data elements serving as location specifiers.  The above theorem
can be extended to a distributed setting where different peers
send around their input data to their peers.  The receiving peer
treats these messages as EDB facts.  This works without
coordination since the program is monotone in the EDB relations.
More generally, it seems one can define a syntactic class of
``oblivious'' Dedalus programs in analogy to our notion of
oblivious transducers.  The restriction would amount
to disallowing joins on location specifiers.

\section{Conclusion} \label{secconcl}

Encouraged by Hellerstein
\cite{hellerstein_datalog,hellerstein_declimp}, we have tried in
this paper to formalize and prove the CALM conjecture.
We do not claim that our approach is the only one that works.
Yet we believe
our approach is natural because it is firmly grounded in previous
database theory practice, and delivers solid results.

Much further work is possible; we list a few obvious topics:
\begin{itemize}
\item
Look at Hellerstein's other conjectures.
\item
Investigate the expressiveness of variations or
extensions of the basic distributed computation model
presented here.
\item
Understand the exact expressive power of the Dedalus
language, as well as the automated verification of Dedalus
programs.
\item
Identify special cases where
essential semantic notions such as
monotonicity, consistency, network-topology independence,
coordination-freeness, are decidable.
\end{itemize}

\bibliographystyle{plain}
\bibliography{database}

\end{document}